# Dialectics of antimicrobial peptides I: common mechanisms of offensive and protecting roles of the peptides

Marta V. Volovik[1], Zaret G. Denieva[1], Oleg V. Kondrashov[1], Sergey A. Akimov[1], Oleg V. Batishchev[1].

[1]Laboratory of Bioelectrochemistry, A.N. Frumkin Institute of Physical Chemistry and Electrochemistry, Russian Academy of Sciences, 31/4 Leninskiy prospekt, Moscow 119071, Russia

Correspondence: olegbati@gmail.com (O.V.B.); akimov_sergey@mail.ru (S.A.A.)

Antimicrobial peptides (AMPs) have intrigued researchers for decades due to the contradiction between their high potential against resistant bacteria and the inability to find a structure-function relationship for the development of an effective and non-toxic agent. In the present study and the companion paper [Phys. Rev. E (2024)], we performed a comprehensive experimental and theoretical analysis of various aspects of AMP-membrane interactions and AMP-induced pore formation. Using the well-known melittin and magainin as examples, we showed, using patch-clamp and fluorescence measurements, that these peptides, even at nanomolar concentrations, modify the membrane by making it permeable to protons (and, possibly, water), but not to ions, and protect the membrane from large pore formation after subsequent addition of 20-fold higher concentrations of AMPs. This protective effect is independent of the membrane side (or both sides) of the peptide addition and is determined by the peptide-induced deformations of the membrane. Peptides create small, $H^+$-permeable pores that incessantly connect the opposing membrane leaflets, allowing translocation of peptides and lipids and thus preventing further generation of large lateral pressure/tension imbalance. At the same time, such an imbalance is a key to the formation of peptide-induced pores at high AMP concentrations, with the main contribution coming from single ion-conducting events rather than stable channel-like structures. Therefore, our results suggest that lowering the AMP concentration, which is a common principle to reduce toxicity, may actually make bacteria resistant to AMP. However, a protective pre-treatment with nanomolar concentrations of peptides may be the key to protect eukaryotic cells from the high concentrations of AMPs.

## I. INTRODUCTION

In the late 1980s - early 1990s, the problem of microbial resistance to antibiotics became a global issue. For instance, as tetracycline, streptomycin, and chloramphenicol were first introduced in the 1950s, resistant strains to those antibiotics made up 0.04 percent, but within twenty years their number had risen to 90% [1]. The natural defense system of a great variety of living organisms was found to produce endogenous peptides, which provide a strong resistance to pathogenetic microorganisms [2–4]. These antimicrobial peptides (AMPs) demonstrate a broad-spectrum activity against viruses [5], bacterial [5–9] and fungal infections [10], in cancer treatment [11], immune modulation [12], and anti-inflammatory processes [13,14]. During the last 5 decades, more than 3000 AMP types [15] have been described and isolated from various organisms including mammalians and even humans, plants, insects, fungi, and bacteria. However, despite all advantages, clinical trials confront challenging aspects to use AMPs as a medical drug. These aspects include limited stability in human body, short half-lives, toxicity and low selectivity. In view of these problems, detailed understanding of the structure-function relationships is crucial to define the selectivity and balance between the toxicity and efficacy of AMPs.

Antimicrobial peptides are amphipathic molecules containing predominantly cationic and hydrophobic amino acid residues that provide interactions with both hydrocarbon chains and polar or charged headgroups of lipids in a membrane. Almost all of the isolated natural peptides (up to 90%) are small molecules comprising up to 100 amino acids [16]. AMPs fold into an α-helical structure or form β-sheets when adsorbed on the membrane surface, as confirmed by both circular dichroism and NMR studies [17]. AMPs bearing positive charge have a strong affinity to anionic bacterial membranes, but also target zwitterionic lipids as well. In low concentrations, α-helical AMPs adsorb onto the lipid bilayer approximately parallel to the membrane plane, in so-called S-configuration [18–21]. The hydrophobic side of the peptide immerses into one of the membrane leaflets and disturbs the neighboring lipids. As the AMP concentration reaches some threshold value, one can detect the formation of pores in the membrane [22–24]. It is commonly believed that pores result from the cooperative change of the AMP orientation from parallel to perpendicular with regard to the membrane plane (I-configuration) [19,25,26]. Antimicrobial peptides produce pores of different structure, size and lifetime depending on their chemical structure, membrane lipid composition, temperature, and peptide to lipid (P/L) ratio. On the other hand, this critical surface concentration of peptides weakly depends on their chemical structure, and for most AMPs lies in the P/L range of 1/25-1/100 [27,28].

Generally, there are two models of pores formed by AMPs: barrel-stave model [29,30], which is characterized by the pore lumen completely lined with AMP molecules; and toroidal model [24,27] of a mixed lipid-peptide pore edge. Toroidal pore formation is predominantly associated with the peptides in the I-configuration; at the same time, in [31–33] the authors suggest that AMPs in the toroidal pore are oriented parallel to the membrane plane. Such a variation of the toroidal model is called the sinking raft model. At very high concentrations, AMPs can act as a detergent [34], solubilizing the membrane, in accordance with the so-called carpet model of membrane disintegration by peptides. However, all of these models do not explicitly consider the trajectory AMP transition from the S-configuration to the I-configuration. Peptide reorientation most probably occurs through some intermediate states that are too short-living to be observed experimentally. Due to the peptide amphipathic nature, even in the S-configuration, AMPs immersed into the lipid monolayer should deform the membrane [35]. Recently, we have shown that for P/L ratio of about 1/100 the whole membrane surface should be completely deformed by AMP molecules [36]. Therefore, for the toroidal model, the threshold number of AMPs at the membrane to form pores may not be associated with the formation of any cooperative peptide structure of a certain stoichiometry, but may be determined by the threshold in the probability of intense appearance of lipidic pores.

Melittin and magainin seem to be the two most extensively studied antimicrobial peptides. Melittin is obtained from the venom of European honeybee *Apis mellifera*, while magainin belongs to the amphibian peptide family isolated from the frog skin *Xenopus Laevis* [6,37]. The family includes magainin-I and magainin-II, two closely related peptides with the difference in two amino acid residues. In our study we consider only magainin-I and omit the number for simplicity. Both peptides are α-helical when adsorbed onto a lipid membrane, linear, cationic, and have a length of about 4 nm. Melittin consists of two α-helices joined by the proline loop, comprises 26 amino acids, and bears a net positive charge of +6e, at physiological pH [38,39]. Magainin comprises one helix of 23 amino acids and has a smaller charge of +5e [27,40,41]. In low concentration, melittin interacts with the membrane as a monomer [42]. Meanwhile, computational, NMR and fluorescence studies show that magainin binds to the lipid bilayer as a dimer [43,44]. With an increase in concentration both peptides demonstrate cooperative self-association followed by pore formation [39,44–47]. Therefore, these peptides could be the most representative examples of amphipathic AMPs forming toroidal lipid-peptide pores.

A large variety of experiments with giant and large unilamellar vesicles shows that peptide-induced transient pores yield a burst of leakage events in the first few minutes after peptide addition [28,39,48,49]. At a low peptide to lipid ratio (P/L< 1/200), leakage slows down or stops completely before releasing all the dye. The transient activity is also confirmed by a recent single-cell fluorescence microscopy study on live *Escherichia coli*, where the authors have observed membrane re-sealing in the presence of melittin [50]. In the case of magainin, at high P/L ratio (1/20 – 1/100), the transient pore-forming activity may proceed to a steady-state with the appearance of stable pores [28]. The same is reported for melittin: at high enough concentrations it forms stable pores that are permeable to glucose and other large molecules [51,52]. Thus, the lifetime and size of the AMP-induced pores can be modulated by the P/L ratio. However, a transient permeabilization is observed over the entire range of AMP concentrations. Permeabilization can be coupled to the peptide's translocation across the membrane [50,53,54]. A broad spectrum of conducting events, induced by melittin and magainin, are classified into three main types according to the description of the typical shapes of patch-clamp signals: spike, multi-level and square-top [55]. The spike signal is a sharp jump of the conductance with an immediate drop down to the baseline in less than 20 ms. The spike signal is considered as a short instantaneous fluctuation corresponding to minor membrane perturbations by one or several peptide molecules [56]. Multi-level shape represents a complex set of a broad range of discrete signals indicating transient pores dynamically fluctuating in their lifetime and size. Square-top signals have a step-like shape with a discrete change of the conductance and reflect a steady-state-like membrane permeabilization. In [56], the authors suggest that square-top and multi-level signals are related to the formation of pores described by the barrel-stave or toroidal models, respectively. However, while all types of the signals have been observed for both melittin and magainin, none of them form barrel-stave pores, in contrast to alamethicin [47]. Thus, a single type of AMP can cover the whole possible range of conducting events and types of peptide-induced pores that makes questionable a possible relation between the AMP chemical structure and the architecture of formed pores.

Marks et al. has shown that amphipathic peptides can spontaneously cross the membrane without the formation of detectable pores [57]. Wheaten et al. has also shown peptide translocation without dye leakage using confocal fluorescence microscopy on GUVs [58]. Surprisingly, peptide translocation to the opposing side of the membrane and its redistribution between membrane leaflets presumably inhibits further pore formation. This hypothesis is based on the earlier observations that melittin [59] and magainin [60] permeabilize the vesicles only for a limited period of time: the total leakage of the vesicles can occur at high AMP concentration, but there is no longer high permeability of the membrane after the initial burst of leakage. A similar effect was observed for the amphipathic fusion peptide of influenza hemagglutinin [61]; the leakage of the GUV membrane with respect to calcein stopped in less than 20 minutes. Although this phenomenon has been extensively investigated, there is still a gap in understanding the mechanism of membrane permeabilization by AMPs and further inhibition of their pore forming ability. Later Matsuzaki et al. [44] have suggested that the peptide density at the outer leaflet of the membrane decreases due to the AMP translocation that blocks further pore formation. On the other hand, in experiments AMPs are usually present in an excessive amount, and it is not clear why the translocated molecules cannot be substituted by fast adsorbing from the water solution ones AMPs. Thus, the inhibition mechanisms of pore formation remain an open question.

In the current work we used melittin and magainin to experimentally clarify the AMP-induced pore formation and inhibition mechanisms. Using patch-clamp technique and planar lipid membranes, we demonstrated the detailed nature of ion-conducting pores formed by these peptides and the corresponding occurrence and duration of conducting events. We found conditions for inhibition of ion-conducting pore formation by pre-adsorbed AMPs, as well as the dependence of this process on peptide molecular structure, concentration, and its relation to the type of conducting events. These experiments were independently supported by measuring the calcein leakage from GUVs in the presence of melittin and magainin. In the companion theoretical paper [62], we used the theory of membrane elastic deformations to manifest the nature of all types of these ion-conducting events and the physical mechanisms of their appearance. We showed that the adsorption of the AMP onto a single membrane leaflet yields a lateral pressure and tension in the opposing membrane leaflets that results in the intensive formation of purely lipidic pores. The size and structure of these pores are dictated by the value of the pressure/tension imbalance, which is determined by the area introduced by a single adsorbing AMP molecule and its surface concentration. In addition, large peptide-lipid pores can form at high enough surface concentration of AMP even if it is adsorbed symmetrically onto both membrane monolayers. The conditions of large peptide-lipid pore formation are determined in the companion paper [62].

We performed fluorescent experiments with GUVs to analyze the peptide-induced membrane permeabilization at very low (nanomolar) concentrations. We found that in concentrations less than required for the formation of ion-conducting pores, AMP induces the formation of small pores, permeable only to protons (hereinafter, $H^+$-pores). In the companion theoretical paper [62], we predicted that AMPs deform the membrane in a way favoring an appearance of small lipid-peptide $H^+$-pores impermeable to water-soluble fluorescent dyes and ions, but permeable to protons (and, possibly, water). The $H^+$-pores incessantly connect two membrane leaflets, allowing them to exchange by lipids and peptides. Such $H^+$-pores can relieve a sharp difference in the lateral pressure/tension between the membrane leaflets, thus protecting membranes from further formation of large (leaky) pores even upon a subsequent addition of 20-fold higher concentration of AMPs. Therefore, in a set of two companion papers we both experimentally and theoretically proposed an original new (as far as we know), and comprehensive mechanisms of AMP-induced pore formation and inhibition.

## II. MATERIALS AND METHODS

Materials

     1,2-dioleoyl-*sn*-glycero-3-phosphocholine (DOPC), 1,2-dioleoyl-*sn*-glycero-3-phospho-rac-(1-glycerol) sodium salt (DOPG), 1,2-dioleoyl-*sn*-glycero-3-phosphoethanolamine (DOPE), and 1',3'-bis[1,2-dioleoyl-sn-glycero-3-phospho]-glycerol (TOCL) were purchased from Avanti Polar Lipids (Alabaster, AL, USA). Lipid solutions were prepared using chloroform, squalane, octane, and decane (all from Sigma Aldrich, Saint-Louis, MO, USA, purity >99%). Anhydrous glucose and sucrose (both from SERVA, Heidelberg, Germany), KCl (Sigma-Aldrich, Saint-Louis, MO, USA), HEPES (Helicon, Moscow, Russia), bovine serum albumin (BSA) (Sigma-Aldrich, Saint-Louis, MO, USA, purity >98%), melittin (Sigma-Aldrich, Saint-Louis, MO, USA, purity >97%, HPLC), magainin-I (Sigma-Aldrich, Saint-Louis, MO, USA, purity >97%, HPLC), and calcein (Sigma-Aldrich, Saint-Louis, MO, USA), were used without further purification.

Methods

Formation of planar bilayer lipid membranes (BLMs)

     Lipid stock solutions of 10 mg/ml lipid in chloroform were mixed in a 1.5-ml Eppendorf vial to obtain the composition of DOPC:DOPG:DOPE = 60:20:20 by mol%. The solvent was removed by drying under an argon stream for 30 min, and then the mixture was dissolved in squalane at a concentration of 20 mg/ml. Bilayer lipid membranes were formed by the Muller-Rudin technique [63] on the 100 µm x 100 µm cooper mesh placed in a Petri dish. The mesh was preliminarily covered with 0.6 µl of the lipid solution in 1:1 (v/v) mixture of octane and decane with a concentration of 10 mg/ml, and dried under an argon stream to form a meniscus. The Petri dish was filled with working buffer, WB (100 mM KCl, 10 mM HEPES, pH 7.5) in Milli-Q (MilliPore, Direct-Q 3UV system, Burlington, MA, USA) water. The mesh was painted with the lipid solution in squalane, and BLMs were formed spontaneously.

Patch-clamp measurements

     Measurements of the electrical current through the membrane were performed using a patch-clamp amplifier (HEKA EPC-8, Lambrecht, Germany) in voltage-clamp mode with a pair of Ag/AgCl electrodes. The ground electrode was placed in the WB in a Petri dish, while the

measuring one was placed in the glass micropipette filled with the same working buffer. After establishing a tight contact between the pipette and the BLM, a constant voltage of +100 mV was applied to the electrodes, and the current through the membrane was measured.

The patch-clamp experiments were divided into two stages. At the first stage, to monitor the pore formation by AMPs, the micropipette was filled with the peptides dissolved in the working buffer in the high concentration of 700 nM for melittin and 800 nM for magainin. In the second stage, the same experiment was conducted with the peptides added into the working buffer solution in a Petri dish in the low concentrations of 17.5 nM and 35 nM for melittin or 20 nM and 40 nM for magainin. In each experiment the membrane conductance was recorded for 15 minutes, and a total number of 30 experiments was performed for every condition of measurements.

To analyze the pore-forming activity of melittin and magainin over the course of the experiment, we calculated the number of experiments with conducting event for each minute of the recording and divided them by the total number of experiments. We use the definition of conducting events from [55]. The results of the calculations were presented as the relative occurrence of conducting events. The percentage of conducting event types (spike, multi-level or square-top) was estimated as the number of signals of the corresponding type during the total time of all experiments divided by the total amount of all signal types and multiplied by 100 %. A conducting event was considered reliable if its value was greater than twice the average noise level.

As a control, we conducted experiments with bovine serum albumin (BSA), which is known for its ability to bind negatively charged membranes [64]. In the first set of control experiments, the current through the membrane was measured with the micropipette filled with BSA in the WB in the high concentration of 800 nM. The concentration of BSA in a Petri dish was either 0, 20 nM, 40 nM, or 100 nM. In the second set of control experiments, magainin was added to the micropipette in the same working buffer in a high concentration of 800 nM, while the Petri dish was filled with BSA in a low concentration of 40 nM.

Formation of giant unilamellar vesicles (GUVs)

GUVs were prepared by the method of giant suspended bilayers [65], slightly modified as follows to prepare vesicles more rapidly as compared to the classical method of electroformation [66]. Lipid stock solutions of 10 mg/ml lipid in chloroform were mixed in a 1.5-ml Eppendorf vial to obtain the composition DOPC:DOPG:DOPE = 60:20:20 by mol.% and DOPE:DOPG:TOCL = 70:20:10 mol.%. The solvent was removed by drying under an argon stream for 30 minutes. Multilamellar lipid vesicles (MLVs) were formed by hydration of the lipid film in 200 μl of Milli-Q water (MilliPore, Direct-Q 3UV system, Burlington, MA, USA) with a final concentration of lipid in water of 1 mg/ml. MLV solution was stored at 4 °C and used in less than three days.

MLV solution in 8–10 equal drops of 3 μl each was placed on a clean Teflon surface. Each drop was gently touched with a pipette tip containing 8 μl of 212-300 μm glass beads (Sigma Aldrich, Saint-Louis, MO, USA). The beads were preliminarily washed three times with Milli-Q water and stored in it. Bead-MLV drops were dried in vacuum for 40 minutes for complete water evaporation. Dry beads covered by a lipid film were stored at 4 °C for less than 24 h.

A 10 μl plastic pipette tip was cut from the bottom to approximately 2/3 of its original size. The cut tip was used to take 6 μl of the hydration buffer (HB) containing 500 mM sucrose, 1 mM HEPES, pH 7.5, and 0.16 mM of calcein. Then the tip was carefully detached from the pipette and held in a vertical position all the time. A small portion of lipid-covered beads from one of the dried drops was picked up and deposited into the tip from the top. The tip was then carefully introduced into a clamp and incubated at 45 °C for 15 minutes.

The beads covered by the hydrated lipid film were transferred into a microcentrifuge tube containing 30 μl of the hydration buffer. The plastic tip containing hydrated beads was carefully withdrawn from the clamp and the end of the tip was put into tight contact with the interface of the buffer followed by sedimentation of the beads to the bottom of the tube by gravity. The beads were incubated in the buffer for 20 minutes at 45 °C with minimal agitation to obtain GUVs. Next, GUVs were separated from the beads by gentle pipetting with a wide tip. The vesicles produced this way were mostly spherical and unilamellar, with diameters ranging from 5 to 50 μm.

Fluorescence microscopy on GUVs

The interactions between antimicrobial peptides and GUVs were investigated using the Ti Eclipse fluorescence microscope (Nikon, Tokyo, Japan) with a 60x magnification objective using the single GUV method [67]. A Petri dish with a glass bottom pre-treated with BSA in a concentration of 1 mg/ml was filled with the WB containing 300 mM glucose. 10 μl of the GUV solution was transferred to the microscopy chamber for further observation. All experiments were performed at room temperature and in isosmotic conditions. The osmolarities of the buffer solutions were controlled by an osmometer (Osmomat 030, Gonotec GmbH, Germany). Vesicles were selected based on the criteria that they were attached to the bottom of the experimental chamber, were unilamellar, and had no visible protuberances.

Analysis of GUV diameter before and after addition of peptides (with and without pre-adsorption of the peptide in a low non-pore-forming concentration) was performed by phase-contrast microscopy. The initial GUV diameter ($D_{initial}$) was calculated before the introduction of the peptide. The final GUV diameter ($D_{final}$) was determined 10 minutes after the start of the peptide addition in a pore-forming concentration. Diameters were checked between at least three cross-sections, and only GUVs with deviation of diameters less than 1% were taken for further analysis. The change in the diameter ($\Delta D$) was calculated as follows:

$$\Delta D\ (\%) = \left(\frac{D_{final}}{D_{initial}} - 1\right) \times 100\%$$

Calcein leakage assay

The process of pore formation by melittin or magainin was investigated by the change of GUV fluorescence intensity due to calcein leakage. Each peptide solution in the WB containing 300 mM glucose was introduced through a glass micropipette with a 1 μm tip diameter in the vicinity of a target GUV containing the HB. The distance between the GUV and the tip of the micropipette was 50 μm, and a small positive pressure was applied to induce the peptide solution flow from the pipette. Experiments were performed in isosmotic conditions between the GUV interior and the external solution (~525 mOsm) and at room temperature. A complete (100 %) leakage of calcein was performed by the addition of Triton X-100 in a final concentration of 0.07 % (v/v). Calcein leakage was monitored using a camera (Teledyne Lumenera, Ottawa, Canada). Images were saved in InfinityCapture software and processed using ImageJ [68]. The average intensity per GUV was estimated. The release of fluorescent dye from the GUV was measured as follows:

$$leakage\ (\%) = \frac{I_m - I_0}{I_{100} - I_0} \times 100\ \%,$$

where $I_m$ is the measured fluorescence intensity, $I_0$ and $I_{100}$ are the fluorescence intensities at the time zero and after the addition of Triton X-100, respectively. These intensities were taken as an integral value inside the GUV. Calcein leakage experiments were performed several times using different GUV samples on different days. The results presented here are the average of the experiments, ± standard deviation (SD). Each of them was obtained from 3-5 independent preparations of GUV, and 10 single GUVs in each preparation were analyzed.

Proton leakage assay

GUVs loaded with fluorescent dye were used to investigate the ability of AMPs to produce small pores permeable only to protons. Calcein was used as a fluorescent pH-sensitive marker, whose intensity at pH 5.0 is always less than at pH 7.5 [69]. GUVs were prepared as described above in HB without HEPES, at pH 5.0, with 0.16 mM calcein. The acidic pH inside GUVs was achieved by adding 1 M HCl. HEPES was not used in the hydration buffer to exclude the effect of buffering on the proton leakage. Peptides in non-pore-forming concentrations were introduced through a micropipette in the vicinity of a GUV in the WB containing 300 mM of glucose. The experiments were performed in isosmotic conditions between GUV interior and external solution (~511 mOsm) and at room temperature. The flow of protons out of the vesicle was estimated using densiometric graphs plotted over the GUV diameter. The gray values were defined as the difference between the average brightness level per GUV and the average brightness of the vesicle environment and normalized in the following order. The gray value for each pixel of the GUV diameter in the presence of the peptide was divided by the gray value of the corresponding pixel for the intact GUVs in the absence of peptides at the initial time point (0 s). The diameter of the GUV was approximately 200-250 pixels. Experiments for each peptide were performed on different vesicle samples. For each experiment, there were 3-5 independent preparations of GUVs, and 10 single GUVs in each preparation were analyzed.

**III. RESULTS**

Electrical measurements reveal three types of peptide-induced pores

To analyze in details the process of pore formation by melittin and magainin at different concentrations, we recorded the current through the lipid bilayer using the patch-clamp technique. It is known for decades that both peptides are able to form pores in bacterial membranes [6,37]. Moreover, recently we have shown that differences in their membrane affinity can be described in terms of electrostatic interactions by the Gouy-Chapman-Stern model [70]. Therefore, to assess common principles of membrane activity of these α-helical AMPs, we decided to take a lipid composition DOPC:DOPG:DOPE in a molar ratio of 60:20:20, which allows us to obtain both stable planar BLM and GUV that is used in many studies of AMP membrane activity in different ratios of the components, as reviewed in [71]. This composition comprises PG lipids in amounts typical for bacteria and necessary for AMP adsorption, while PC lipids are more typical for mammalian cell membranes.

The micropipette was filled with melittin or magainin solution in the WB and placed in close contact with BLM. For concentration exceeding the threshold ones, in particular, 700 nM for melittin and 800 nM for magainin, we observed ion-conducting pores of different lifetimes and different types of the electrical signal. We obtained that at these concentrations melittin formed pores in 90 % of the experiments, and magainin — in 87 %. These peptide concentrations correspond to a P/L ratio of about 1/25 for melittin and 1/60 for magainin [72–74], respectively, which lies in the defined range of pore-forming AMP surface concentrations of 1/25-1/100 [27,28].

The distributions of conducting events occurrences for melittin and magainin over the course of experiments are shown in Fig. 1. One can see that the pore-forming activity for both peptides decreased sharply in about 5 minutes from the beginning of the experiment and became almost zero at the end of the experiment (15 min). Indeed, in the first several minutes the peptides induced conducting events with a relative occurrence of about 0.6, while at the 15-th minute this occurrence was reduced to 0.01. These results are consistent with the existing data from vesicle leakage experiments, which have shown that a burst of leakage occurs in the first few minutes, followed by a slowing or cessation of leakage [28,44,48].

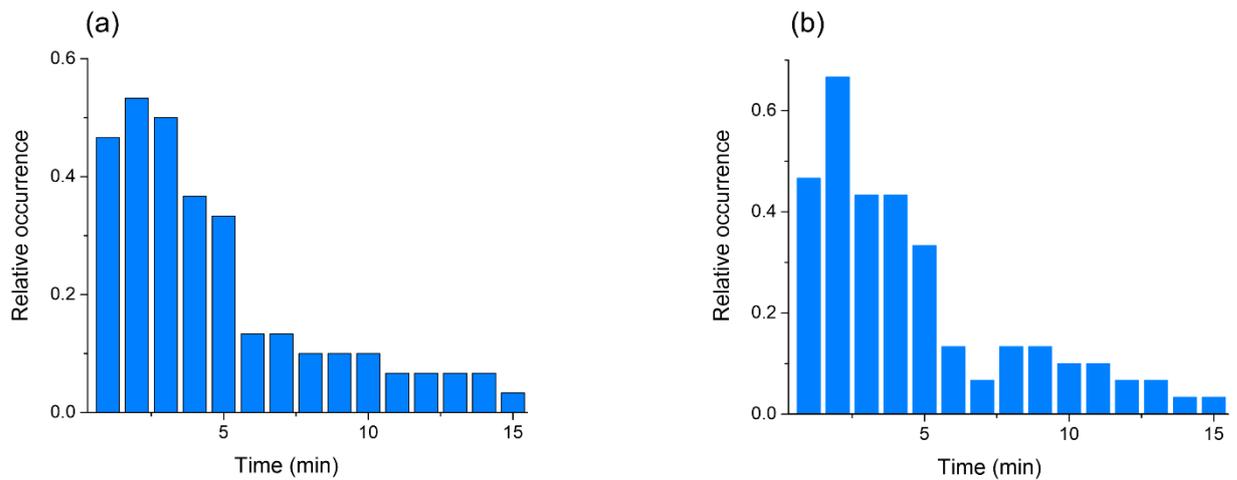

FIG. 1. Histograms of the relative occurrence of conducting events over the course of the experiments (15 min) for 700 nM melittin (a) and 800 nM magainin (b), summarized from 30 experiments for each peptide.

Both peptides, melittin and magainin, demonstrated a broad spectrum of electrical signals with an increase of the membrane conductance and subsequent drop to its initial values. Peak values vary in a range of conductance from ~0.5 nS to ~2.5 nS, which agrees with other electrophysiological studies on melittin [75]. These signals were classified into three main groups according to the description of the typical shapes of patch-clamp signals for AMP-induced pores [55]. Analysis of the conductance traces for melittin and magainin in the concentrations of 700 and 800 nM, respectively, revealed no significant difference in all three types of signals between two peptides. Figure 2 represents the percentage of spike, multi-level and square-top signals in 30 experiments for each of the peptides.

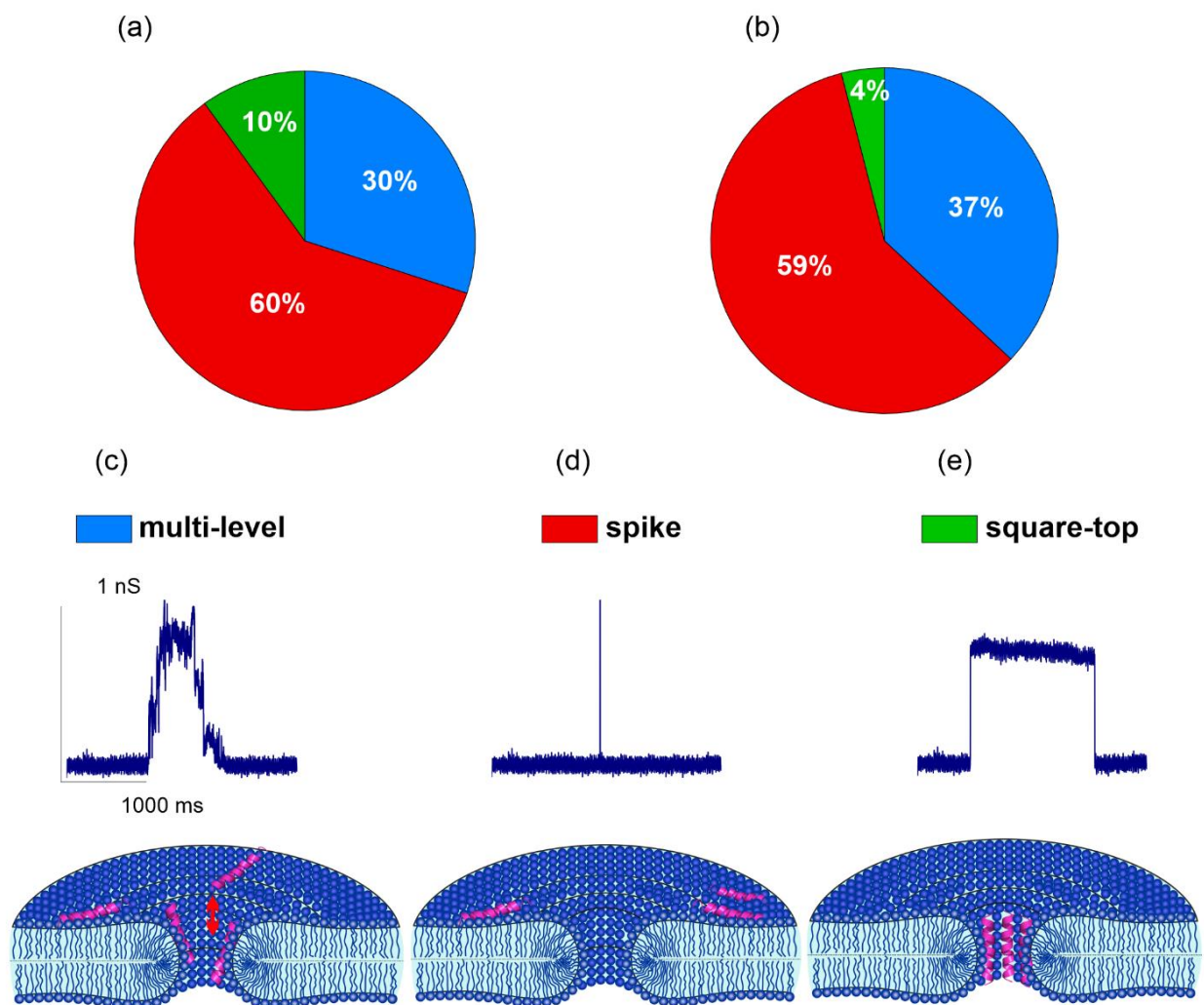

FIG. 2. Diagrams of the percentage of spike, multi-level, and square-top signals in 30 experiments for 700 nM melittin (a) and 800 nM magainin (b). Total number of conductance events was 299 for melittin and 419 for magainin. Typical examples of the three conductance signals and its structural equivalents (c-e). Multi-level conductance traces most probably represent peptide-lipid pores with the peptides dynamically fluctuating in them, so that the tilted peptide molecules can stand at the edge of the pore or escape from it (c). Spike conductance corresponds to the lipidic pore, formed by the adsorbed peptides that generate the imbalance in the lateral pressure/tension between the membrane leaflets (d). Square-top signals are some long-living stable structures formed by lipids and several peptide molecules, apparently standing in the I- configuration at the pore rim. This configuration rarely appeared in the system (e).

Melittin and magainin demonstrated similar values of about 30-37% in the percentage of multi-level signals and 59-60 % in the percentage of spike signals. Square-top signal appeared quite rarely for both peptides. Spike conductance represents the most abundant type of signals: 60% for melittin and 59% for magainin. Duration of each type of the signal was from 1 to 20 ms for spike and up to 10 minutes for multi-level/square-top signal. Our results were in agreement with previous studies which showed 67% of spike signals for melittin [75] at about 1 micromolar concentrations. The percentage of multi-level signals, which occurred in 30% of cases for melittin and 37% for magainin, was also consistent with previous observations [76]. Due to the similarity of erratic conductance, which is considered as a burst of peaks with no apparent structure or distinct shape [55], and multi-level conductance, we considered these two signals, according to [76], as a single type.

In [75], the multi-level signal is associated with toroidal pore model, while the erratic one is considered to correspond to the carpet model. However, the authors note that these two signal types can be exchangeable due to the similarity of the signal distribution and the difficulty in their distinguishing. Considering these aspects, we can emphasize, that the ratio of multi-level conducting events for melittin was close to the ratio of erratic signals (33%) on the DPhPC and DPhPC/POPG bilayers at similar concentrations (around 1 μM) as observed in [75]. The square-top signal represents the minor percentage of conducting events: 10 % for melittin and 4 % for magainin. Our results are consistent with the electrophysiology measurements on the bacterial membranes, which demonstrated 11 % of square-top signals for magainin in the micromolar concentrations [77].

To estimate the minimal size of ion-conducting pores, we analyzed the conductance of spike signals. Values of the pore radius were obtained from the spike amplitudes. The following equation was used to evaluate the pore radius:

$$R = (l + \frac{\pi r}{2})\frac{\rho}{\pi r^2}$$

where $R$ is the electrical resistance of the pore obtained from the experimental data, $l = 4$ nm is the membrane thickness, $\rho = 1.264$ $\Omega^{-1}m^{-1}$ is the resistivity of the buffer solution, $r$ is the pore lumen radius [78].

Calculations imply that the most frequently occurring pore radius was 1.2 nm for melittin and 1.1 nm for magainin (Fig. 3). Note that these distributions cannot be approximated by the Gaussian function because the lower limit of the pore size is in the least determined by the size of the hydrated ion passing through the pore.

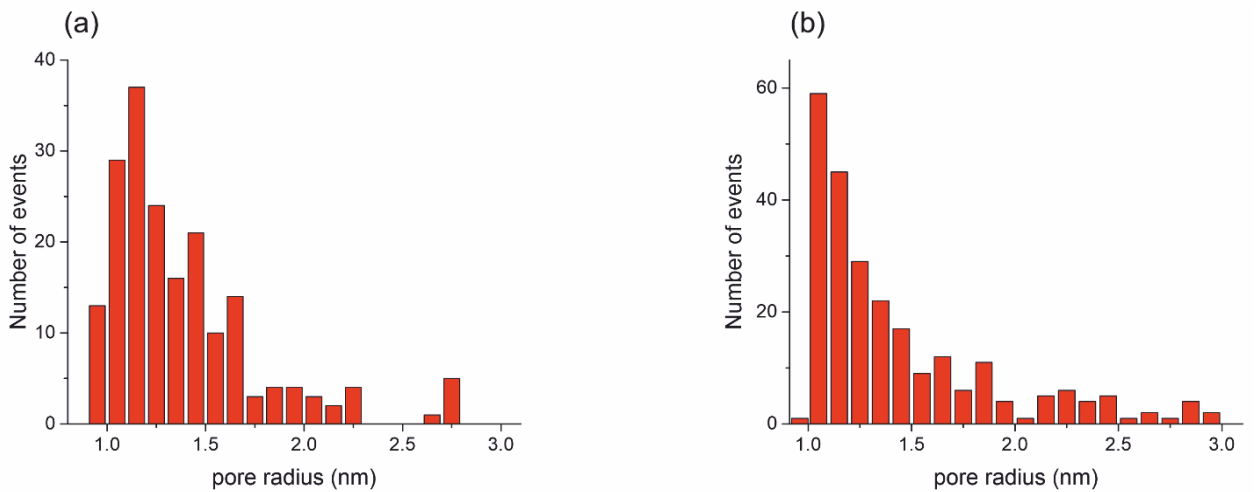

FIG. 3. Histograms of the pore radius distribution across all experiments for 700 nM melittin (a) and 800 nM magainin (b) obtained from the spike amplitudes. The most frequently occurring pore radius is 1.2 nm for melittin (700 nM) and 1.1 nm for magainin (800 nM).

Peptide pre-adsorption on both membrane leaflets inhibits all types of pores

To analyze the pore inhibiting effect of the pre-adsorption of melittin and magainin in low concentrations, we performed the following experiments. A Petri dish was filled with WB and the membranes were formed. Next, the peptide was added to the Petri dish and the concentrations in the solution were 35 nM for melittin or 40 nM for magainin. These values are 20 times lower than the threshold concentrations for the appearance of peptide-induced pores in a membrane. After about 10 minutes, a micropipette containing the solution with the high peptide pore-forming concentration of 700 nM for melittin or 800 nM for magainin, was placed in contact with the BLM. During the experiments, we found that the occurrence and the duration of different signal types significantly decreased in the presence of peptides pre-adsorbed onto the membrane in low concentration. The number of spike, multi-level and square-top signals for melittin and magainin with and without pre-adsorption are shown in Table I, while Table II shows the duration of the respective signals in the corresponding experimental conditions. While we observed a decrease in the initial number of multi-level and square-top signals to 27 % and 24 % for pre-adsorbed 35 nM melittin, and to 28% and 22% for pre-adsorbed 40 nM magainin, there was a sharp drop in the number of spike signals to 4 % for melittin and to 9 % for magainin for the same conditions (Table I). The duration of the spike/multi-level/square-top signal in the presence of these concentrations of peptides in a Petri dish decreased to 5/11/55 % for melittin and to 7/6/44 % for magainin, respectively (Table II).

TABLE I. Number of spike, multi-level, and square-top signals upon addition of 700 nM melittin or 800 nM magainin with or without pre-adsorption of the same peptide in non-pore-forming concentrations. Data were taken from 30 experiments that lasted 15 minutes each for both melittin and magainin.

| Number of signals for melittin | Spike | Multi-level | Square-top | Total |
|---|---|---|---|---|
| Pore formation without pre-adsorption | 181 | 89 | 29 | 299 |
| Pore formation with pre-adsorption of 35 nM | 8 | 24 | 7 | 39 |
| Pore formation with pre-adsorption of 17.5 nM | 24 | 24 | 8 | 56 |
| Ratio of conducting events at 35 nM with/without pre-adsorption, % | 4 | 27 | 24 | 13 |
| Ratio of conducting events at 17.5 nM with/without pre-adsorption, % | 13 | 27 | 28 | 19 |
| Number of signals for magainin | | | | |
| Pore formation without pre-adsorption | 246 | 155 | 18 | 419 |
| Pore formation with pre-adsorption of 40 nM | 21 | 43 | 4 | 68 |
| Pore formation with pre-adsorption of 20 nM | 2 | 20 | 0 | 22 |
| Ratio of conducting events at 40 nM with/without pre-adsorption, % | 9 | 28 | 22 | 16 |
| Ratio of conducting events at 20 nM with/without pre-adsorption, % | 0.8 | 13 | 0 | 5 |

Electrophysiological experiments on planar lipid bilayers and fluorescence microscopy on GUVs have an important physical difference in terms of the "non-local" mechanism of pore formation [62]. Fast one-sided adsorption of the peptides generates an imbalance in the lateral pressure/tension between the membrane leaflets that results in the formation of the lipidic pores. In the case of the electrophysiological experiment, the planar lipid bilayer has constant lateral tension in the lipid monolayer distant from the patch pipette, as it is continuously connected to the reservoir. However, there is almost no tension in the leaflets of the GUVs. Under otherwise equal conditions, adsorption on a monolayer under tension will be higher [79]. This means that at the same bulk concentration of amphipathic peptide, the surface concentration on the membrane will be higher in the patch-clamp conditions compared to the tensionless membrane of GUV. To verify if the lower concentration of pre-adsorbed peptides would reduce the pressure in the upper monolayer, thus, diminish the pore formation by the non-local mechanism, we performed additional experiments with pre-adsorbed peptides in the concentration of 17.5 nM for melittin and 20 nM for magainin. Surprisingly, decrease in the pre-adsorbed amount of peptides twice led to an almost complete inhibition of pore formation in the presence of 800 nM of magainin in both number and duration of conducting events, while for 700 nM melittin only the duration of the signals decreased. These decreases in the number and duration of conducting events suggest that the presence of tension in the planar lipid bilayer requires a lower bulk concentration of peptides to achieve the same surface concentration of pre-adsorbed peptides, providing the pore inhibiting effect, as compared to the tensionless GUVs. The difference in the inhibiting effect for melittin and magainin is considered in details in the companion paper [62]

TABLE II. The total duration of spike, multi-level, and square-top signals upon addition of 700 nM melittin or 800 nM magainin with or without pre-adsorption of the same peptide in non-pore-forming concentrations. Data were taken from 30 experiments of 15 minutes each for both melittin and magainin.

| Duration of signals for melittin | Spike (ms) | Multi-level (ms) | Square-top (ms) |
|---|---|---|---|
| Pore formation without pre-adsorption | 1048 | 2015796 | 1701199 |
| Pore formation with pre-adsorption of 35 nM | 53 | 222057 | 938237 |
| Pore formation with pre-adsorption of 17.5 nM | 85 | 25609 | 46155 |
| Ratio of durations at 35 nM with/without pre-adsorption, % | 5 | 11 | 55 |
| Ratio of durations at 17.5 nM with/without pre-adsorption, % | 8 | 0.01 | 0.03 |
| Duration of signals for magainin | | | |
| Pore formation without pre-adsorption | 1263 | 2985399 | 447366 |
| Pore formation with pre-adsorption of 40 nM | 89 | 191662 | 197815 |
| Pore formation with pre-adsorption of 20 nM | 7 | 5882 | 0 |
| Ratio of durations at 40 nM with/without pre-adsorption, % | 7 | 6 | 44 |
| Ratio of durations at 20 nM with/without pre-adsorption, % | 1 | 0.2 | 0 |

To show that the inhibition of pore formation is a specific feature of AMPs, we conducted three types of control experiments with BSA, which is able to electrostatically bind to the negatively charged lipid membrane [80]. First, we measured the conductance of the membrane patch isolated by the micropipette filled with BSA solution in the concentration of 800 nM. Next, we repeated the same experiments but in the presence of BSA in a concentration of 40 nM in the working buffer solution in a Petri dish. We did not detect any conducting events. Then, we performed experiments with BSA in the concentrations of 20 nM, 40 nM, and 100 nM in a Petri dish and magainin in the concentration of 800 nM in a micropipette to exclude the possible effect of BSA on the inhibition of the pore formation by antimicrobial peptides. As we expected, in the presence of BSA in the Petri dish we observed the same electrical signals induced by AMPs as in the case of absent BSA. Thus, the results of these experiments agreed with the key role of the AMP amphipathic structure in the pore inhibition effect.

Peptides induce calcein leakage from GUVs

    We performed fluorescence microscopy experiments on calcein leakage from GUVs induced by melittin or magainin with the same lipid composition of DOPC:DOPG:DOPE in the molar ratio of 60:20:20. The results obtained are shown in Table III.

TABLE III. Leakage of calcein from GUV comprising DOPC:DOPG:DOPE (60:20:20 mol%) or DOPE:DOPG:TOCL (70:20:10 mol%) induced by melittin in the concentration of 700, 105, 70, 35 nM and by magainin in the concentration of 800, 120, 80, 40 nM.

| Melittin (magainin) concentration (nM) | Calcein leakage (%) | |
| --- | --- | --- |
|  | Melittin | Magainin |
| DOPC:DOPG:DOPE = 60:20:20 mol% | | |
| 700 (800) | 52.7 ± 1.8 | 60.5 ± 13.4 |
| 105 (120) | 30.0 ± 5.4 | 23.8 ± 9.4 |
| 70 (80) | 1.7 ± 1.0 | 1.9 ± 0.6 |
| 35 (40) | 2.2 ± 2.0 | 2.4 ± 1.3 |
| DOPE:DOPG:TOCL = 70:20:10 mol% | | |
| 700 (800) | 55.6 ± 4.4 | 60.7 ± 7.0 |
| 35 (40) | 2.8 ± 0.6 | 5.9 ± 2.3 |

    Adding melittin in the concentration of 700 nM resulted in a leakage of 52.7 ± 1.8 % calcein from the GUV within 600 s, which was detected by observing that the average fluorescence intensity per vesicle decreases [Table III, Fig. 4(a), (e)]. A negligible calcein leakage was caused by the addition of 70 nM and 35 nM melittin to the GUV [Table III, Fig. 4(c), (d), (e)]. Phase-contrast images before and after the peptide addition showed that the shape of the GUV did not change in the course of calcein leakage [Fig. 4(a), 1 and 3]. An average value of leakage and the standard error for each melittin concentration were obtained from 3-10 independent experiments performed with different GUV preparations.

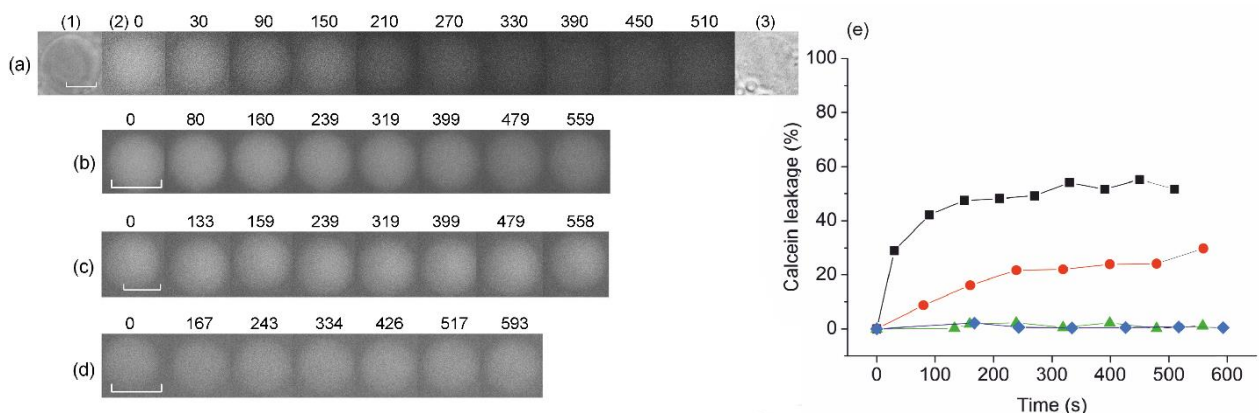

FIG. 4. Calcein leakage from GUV comprising DOPC:DOPG:DOPE (60:20:20 mol.%) induced by melittin. Fluorescence images show calcein release from the GUV after adding melittin in a concentration of (a) 700 nM (1 and 3 show phase-contrast images at 0 s and 510 s); (b) 105 nM; (c) 70 nM; (d) 35 nM. The numbers above each image show the time in seconds passed after we began adding melittin. The scalebar is 10 μm. (e) Typical kinetics of fluorescence intensity changes inside the GUV after the addition of 700 nM (black), 105 nM (red), 70 nM (green), and 35 nM (blue) of melittin. The values were normalized to the average intensity per GUV obtained after the addition of 0.07 % v/v Triton X-100.

For magainin, the introduction of 800 nM of the peptide led to a release of 60.5 ± 13.4 % calcein [Table III, Fig. 5(a), (a)]. A negligible calcein loss was caused by 80 nM and 40 nM magainin [Table III, Fig. 5(c), (d), (e)]. Phase-contrast images before and after the addition of magainin manifested the GUV maintained a spherical shape during the calcein leakage [Fig. 5(a), 1 and 3]. The average value of leakage and the standard error for each magainin concentration were obtained from 4-7 independent experiments performed with different GUV preparations.

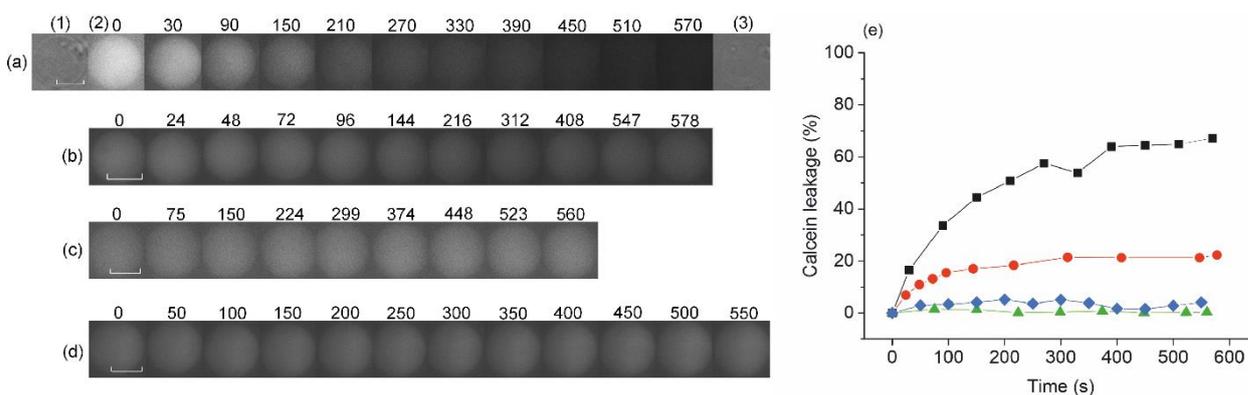

FIG. 5. Calcein leakage from GUV comprising DOPC:DOPG:DOPE (60:20:20 mol.%) induced by magainin. Fluorescence images illustrate the calcein release from the GUV after adding magainin-I in a concentration of (a) 800 nM (1 and 3 show phase-contrast images at 0 s and 570 s); (b) 120 nM; (c) 80 nM; (d) 40 nM. The numbers above each image show the time in seconds passed after we began adding magainin. The scalebar is 10 μm. (e) Typical kinetics of fluorescence intensity changes inside the GUV after adding 800 nM (black), 120 nM (red), 80 nM (green), and 40 nM (blue) of magainin. The values were normalized to the average intensity per GUV obtained after the addition of 0.07% v/v Triton X-100.

Thus, we may conclude that both melittin and magainin formed calcein-leaky pores in the GUV membrane after exceeding some threshold concentration, as it has been reported previously [45,81]. If the peptide was not introduced, stable fluorescence intensity of calcein inside the GUV was observed during the same period of 600 s (data not shown).

To verify that our results would be valid for the typical bacterial lipid composition (by polar headgroup) [71], we also performed additional experiments for GUVs comprising DOPE:DOPG:TOCL = 70:20:10 mol%. The results obtained are shown in Table III.

Adding 700 nM melittin or 800 nM magainin resulted in a leakage of 55.6 ± 4.4% and 60.7 ± 7.0% calcein, respectively [Fig. 6(a), (b), black]. A negligible calcein leakage was caused by adding 35 nM melittin and 40 nM magainin to the GUV [Fig. 6(a), (b), red]. The average value of leakage and the standard error for each melittin concentration were obtained from 3 independent preparations of GUV, and 10 single GUV in each preparation were analyzed.

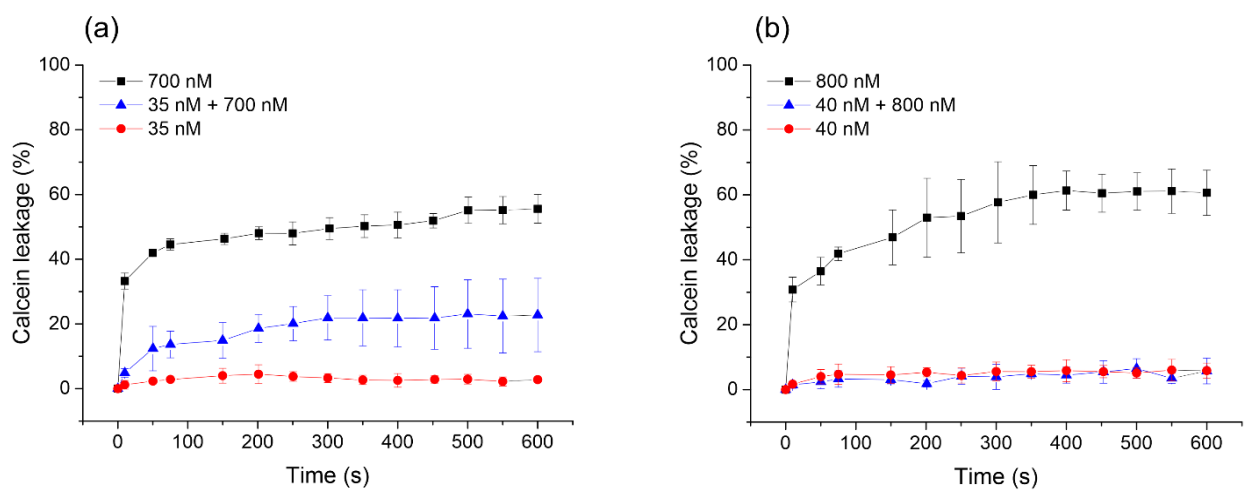

FIG. 6. Kinetics of calcein leakage from GUV comprising DOPE:DOPG:TOCL (70:20:10 mol.%). Addition of 700 nM melittin (black, SD, n = 30); addition of 700 nM melittin after the pre-adsorption of 35 nM melittin solely outer leaflet of the GUV membrane (blue, SD, n = 30); addition of 35 nM melittin (red, SD, n = 30) (a). Addition of 800 nM magainin (black, SD, n = 30); addition of 800 nM melittin after the pre-adsorption of 40 nM magainin solely outer leaflet of the GUV membrane (blue, SD, n = 30); addition of 40 nM magainin (red, SD, n = 30) (b). For the blue curve time zero indicates the moment in which the peptide was added in a high concentration. An average value of leakage and the standard error for each type of experiment were obtained from independent experiments performed with different GUV preparations.

Pre-adsorption of peptides inhibits calcein release from GUVs

Calcein leakage assay manifested that at low concentrations (35 nM for melittin and 40 nM for magainin) these peptides did not induce leaky pores in a GUV membrane. Thus, to study the ability of low concentrations of AMP to block pore formation, we performed the same leakage experiments after pre-adsorption of 35 nM melittin or 40 nM magainin. To verify whether the electrostatic repulsion between adsorbed and incoming peptide molecules is the crucial factor of the pore formation inhibition, we also performed the leakage experiments with the pre-adsorption of 35 nM melittin or 40 nM magainin onto the inner leaflet of the GUV. The obtained results are shown in Table IV.

TABLE IV. Calcein leakage from GUV comprising DOPC:DOPG:DOPE (60:20:20 mol.%) or DOPE:DOPG:TOCL (70:20:10 mol.%) induced by melittin in a concentration of 700 nM and by magainin in a concentration of 800 nM in the presence of pre-adsorbed peptides on the outer, inner membrane leaflet and on both leaflets of the GUV.

| | DOPC:DOPG:DOPE = 60:20:20 mol.% | |
|---|---|---|
| | Calcein leakage (%) | |
| Concentration of pre-adsorption of melittin (magainin) (nM) | Subsequent addition of 700 nM melittin | Subsequent addition of 800 nM magainin |
| Outside the GUV 35 (40) | 23.4 ± 8.7 | 29.8 ± 13.3 |
| Inside the GUV 35 (40) | 24.6 ± 5.9 | 34.1 ± 5.8 |
| Both membrane leaflets 35 (40) | 20.6 ± 2.2 | 1.2 ± 0.7 |
| Outside the GUV 70 (80) | 25.3 ± 2.7 | 1.1 ± 0.5 |
| DOPE:DOPG:TOCL = 70:20:10 mol.% | | |
| Outside the GUV 35 (40) nM | 22.8 ± 11.3% | 5.7 ± 4.0% |

First, 35 nM of melittin were added both inside the GUV by a microinjection and in a vicinity of the GUV from the outside, and incubated for 10 minutes. Piercing of the GUV membrane by the injection pipette led to about 1.1 ± 0.1 % loss of calcein. Subsequent addition of 700 nM melittin [Table IV, Fig. 7(a), black] caused more than twice lower leakage than in the absence of pre-adsorbed 35 nM melittin [Fig. 4(e)]. Second, the effect was investigated on the GUV membrane with melittin in 35 nM concentration applied solely outside or inside the GUV. Only 23.4 ± 8.7 % calcein leakage was caused by adding 700 nM melittin after pre-adsorption of 35 nM melittin from the outside of the GUV [Fig. 7(a), red]. Adding 700 nM melittin after microinjection of 35 nM of the peptide inside the GUV induced the same calcein leakage [Table IV, Fig. 7(a), blue]. Thus, adding simultaneously 35 nM melittin to both membrane leaflets inhibited calcein release from the GUV upon applying 700 nM of the peptide to the same extent as the one-side addition from the outside of the vesicle and from the inside too. To be sure that 10 minutes incubation was enough to reach the equilibrium adsorption of 35 nM melittin onto the GUV membrane, we repeated the experiments with 30 minutes incubation before adding 700 nM melittin. There was no significant difference comparing to 10 minutes incubation results, and calcein leakage was 21.3 ± 1.4 % within 690 s (data not shown). Thus, we concluded that the binding of melittin to the GUV membrane reached equilibrium within 10 minutes.

When the pre-adsorption concentration of melittin was increased up to 70 nM outside the GUV and then 700 nM were adsorbed, a calcein leakage up to 25.3 ± 2.7 % was obtained [Table IV, Fig. 7(a), green]. Together, these experiments indicate that the pre-adsorption of melittin in small, non-pore-forming concentrations onto one or both membrane leaflets can hinder pore formation even for more than one order of magnitude higher concentrations of the peptide applied upon.

Adding 800 nM magainin caused only 1.2 ± 0.7 % calcein leakage after 40 nM of the peptide were pre-adsorbed at both membrane leaflets (Table IV). Adsorbing 40 nM magainin onto the outer or inner leaflet of the GUV led to a partial leakage blocking [Fig. 7(b)]. The case of adsorbing 40 nM magainin on the GUV prior to adding 800 nM magainin solution was also monitored. Calcein leakage reached 30.0 ± 9.7 % within 575 s (data not shown). When the pre-adsorption concentration of magainin was increased up to 80 nM outside the GUV followed by adsorbing 800 nM of the peptide, 1.1 ± 0.5 % calcein leakage was obtained [Table IV, Fig. 7(b), green]. Thus, pre-adsorbing magainin in a low non-pore-forming concentration onto one monolayer of the membrane can inhibit pore formation, and the pre-adsorption onto both membrane leaflets can almost completely prevent further appearance of pores upon adding 10-20 times higher concentrations of the peptide.

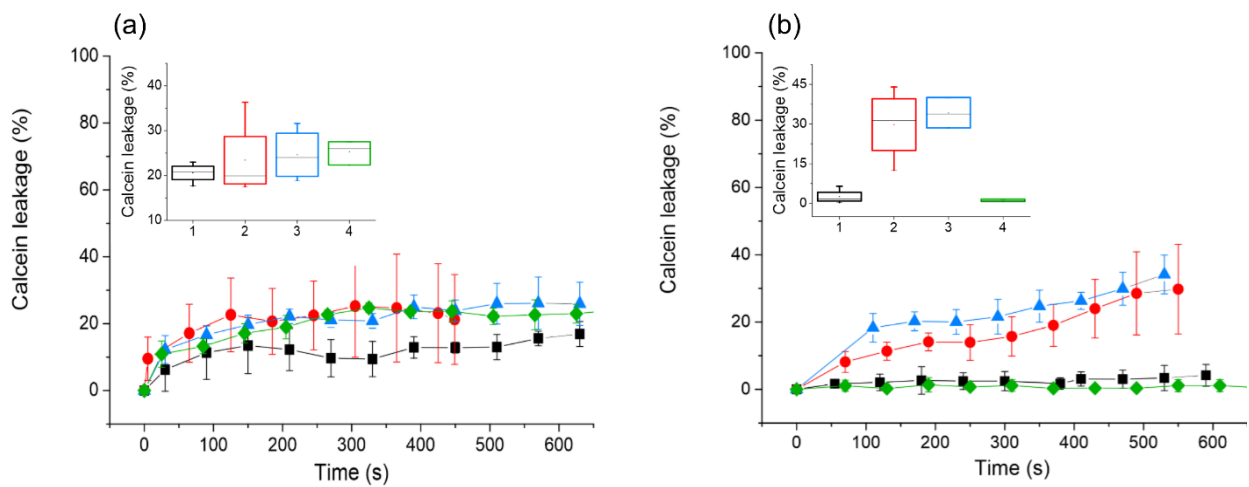

FIG. 7. Kinetics of calcein leakage from GUV comprising DOPC:DOPG:DOPE (60:20:20 mol.%) after pre-treatment with low concentrations of the peptides. Time zero indicates the moment the peptide in a high concentration is added. Addition of 700 nM melittin after the pre-adsorption of: 35 nM melittin onto both (black, SD, n = 40), solely outer (red, SD, n = 40), and solely inner (blue, SD, n = 40) leaflets of the GUV membrane; 70 nM melittin onto solely outer (green, SD, n = 30) leaflet of the GUV membrane (a). The inset shows the distribution of the calcein release value for each listed type of melittin addition (1 to 4 for black, red, blue, green curves, respectively). Addition of 800 nM magainin after the pre-adsorption of: 40 nM magainin onto both (black, SD, n = 40), solely outer (red, SD, n = 40), and solely inner (blue, SD, n = 30) leaflets of the GUV membrane; 80 nM magainin onto solely outer (green, SD, n = 30) leaflet of the GUV membrane (b). The inset shows the distribution of the calcein release value for each listed type of magainin addition (1 to 4 for black, red, blue, green curves, respectively). An average value of leakage and the standard error for each type of experiment were obtained from 3-4 independent preparations of GUV, and 10 single GUV in each preparation were analyzed.

Similar experiments were conducted for the GUVs comprising DOPE:DOPG:TOCL = 70:20:10 mol.%. As described previously, 35 nM of melittin or 40 nM of magainin were added outside the GUV by a microinjection in a vicinity of the GUV, and incubated for 10 minutes. Subsequently adding 700 nM melittin or 800 nM magainin caused 22.8 ± 11.3% or 5.7 ± 4.0% calcein leakage in 600 s [Table IV, Fig. 6(a), (b), blue] that is more than twice lower than the leakage in the absence of pre-adsorbed 35 nM melittin [Fig. 6(a), black] and more than ten times lower than the leakage in the absence of pre-adsorbed 40 nM magainin [Fig. 6(b), black].

While for melittin the results on the DOPE:DOPG:TOCL (70:20:10 mol.%) lipid composition are consistent with the previous experiments on the GUV comprising DOPC:DOPG:DOPE (60:20:20 mol.%), for magainin the leakage was lower for the DOPE:DOPG:TOCL (70:20:10 mol.%) lipid composition compared to the DOPC:DOPG:DOPE (60:20:20 mol.%) one. As we have shown previously [70], near-membrane concentration of the charged molecules of melittin or magainin depends on the surface potential of the membrane as

$$C(0) = C_\mathrm{b} \exp\left(-\frac{zF\varphi_s}{RT}\right),$$

where $C_b$ is the bulk concentration of the adsorbing peptide, $z$ is its charge number, $F$ is the Faraday constant, $R$ is the gas constant, $T$ is the absolute temperature, and $\varphi_s$ is the surface potential at the membrane-water interface. For melittin $z$ = 1.9 [73], while for magainin it is 3.7 [74,82].

According to the Gouy-Chapman model, surface potential depends on the surface charge density at the membrane in the case of binary electrolyte solution as:

$$\varphi_s = \frac{2RT}{F} arsinh\left(\frac{\sigma}{\sqrt{8RT\varepsilon\varepsilon_0 C_{el}}}\right),$$

where $\sigma$ is the surface charge density at the membrane, $\varepsilon$ and $\varepsilon_0$ are the dielectric constants in the solution and in vacuum, respectively, $C_{el}$ is the concentration of the electrolyte ions. As both DOPG and TOCL are charged lipids, for DOPE:DOPG:TOCL (70:20:10 mol.%) surface charge density should be higher than for DOPC:DOPG:DOPE (60:20:20 mol.%). For the latter membrane composition in WB conditions $\varphi_s$ should be around −15 mV [70], and for DOPE:DOPG:TOCL (70:20:10 mol.%) it should be approximately −19 mV [83]. Thus, for the same bulk concentration of magainin in the solution, its near-surface concentration should be ~2 times higher at DOPE:DOPG:TOCL (70:20:10 mol.%) membrane compared to the DOPC:DOPG:DOPE (60:20:20 mol.%) one. Thus, for the 40 nM of magainin added outside the GUV from DOPE:DOPG:TOCL (70:20:10 mol.%) we should see the same effect as for 80 nM added to the DOPC:DOPG:DOPE (60:20:20 mol.%) vesicle. This is exactly the result we observed experimentally (see Table IV).

Peptide adsorption increases GUV diameter

To contradict the hypothesis that in the pore-forming concentration peptides just dissolve the membrane by forming peptide-lipid micelles, we investigated the changes in the GUV diameter in the course of fluorescent microscopy experiments. The experimental system of GUVs permits direct optical monitoring of changes in their size (Table V). Both with and without pre-adsorption of 35 nM melittin onto both membrane leaflets, the diameter of the GUV comprising DOPC:DOPG:DOPE (60:20:20 mol.%) increased by ~2.5 % in 10 minutes after adding 700 nM of peptide. In the case of magainin, adding 800 nM of peptide resulted in an approximately 2 % increase of the GUV diameter in 10 min, while pre-adsorbing 40 nM magainin onto both GUV membrane monolayers diminished the effect almost twice (less than 1% of the GUV diameter growth). Pre-adsorbing peptides in non-pore-forming concentrations only onto one membrane leaflet and afterwards adsorbing them in pore-forming concentrations also increased the GUV diameter approximately by 2 % and 1 % for melittin and magainin, respectively. These results suggest that the increase in GUV diameter for melittin occurred independently of peptide pre-adsorption, consistent with the previous study [84], whereas for magainin, peptide pre-adsorption reduced the diameter change two-fold.

The diameter of the GUV comprising DOPE:DOPG:TOCL (70:20:10 mol.%) increased by 2.7 ± 0.9 % after adding 700 nM of melittin and 3.3 ± 0.4% after adding 800 nM of magainin in 10 min. Pre-adsorbing peptides in non-pore-forming concentrations onto one membrane leaflet and afterwards adsorbing them in pore-forming concentrations also increased the GUV diameter by 2.1 ± 0.6% for melittin and 2.2 ± 0.5% for magainin (Table V).

The results on the DOPE:DOPG:TOCL (70:20:10 mol.%) are consistent with the previous experiments on the GUV comprising DOPC:DOPG:DOPE (60:20:20 mol.%). The increase in the GUV diameter for both peptides means that the peptides do not dissolve the membrane or induce the formation of the peptide-lipid micelles.

TABLE V. Average increase in the GUV diameter (in %) after adding pore-forming concentrations of melittin or magainin. Pre-adsorption concentrations were 35 nM for melittin and 40 nM for magainin, respectively.

| Diameter increase (%) | DOPC:DOPG:DOPE = 60:20:20 mol.% | |
|---|---|---|
| | Peptide | |
| | melittin | magainin |
| Without pre-adsorption | 2.6 ± 0.5 | 2.4 ± 0.9 |
| Pre-adsorption onto both membrane leaflets | 2.4 ± 0.3 | 0.9 ± 0.4 |
| Pre-adsorption from inside the GUV | 2.5 ± 0.5 | 1.8 ± 0.5 |
| Pre-adsorption from outside the GUV | 2.3 ± 0.5 | 1.7 ± 0.5 |
| Pre-adsorption from outside the GUV (70 nM for melittin and 80 nM for magainin) | 2.5 ± 0.6 | 1.6 ± 0.2 |
| DOPE:DOPG:TOCL = 70:20:10 mol.% | | |
| Without pre-adsorption | 2.7 ± 0.9 | 3.3 ± 0.4 |
| pre-adsorption from outside the GUV | 2.1 ± 0.6 | 2.2 ± 0.5 |

Peptides induce proton leakage in non-pore-forming concentrations

It is reported that low concentration of AMPs may induce proton transport [85]. Therefore, we used calcein as a pH-sensitive fluorescent probe in the range of pH 4.0-8.0 in order to evaluate the possible translocation of protons through the GUV membrane in the presence of peptides. Time series of densiometric graphs of the GUV fluorescent microscopy images were plotted to detect the flow of protons out of the vesicle due to the action of peptides. Calcein fluorescence intensity remained stable in a gradient of pH from 5.0 to 7.5 between internal and external solutions of an intact GUV (at absent peptides) over the entire time of the experiments. Addition of 70 nM melittin [Fig. 8(a)] and 80 nM magainin [Fig. 8(b)] caused an increase in the calcein fluorescence intensity rapidly after the start of the peptide addition, and reached a steady-state value in approximately 600 s. The increase in the fluorescence intensity was consistent with a pH increase inside the GUV. No significant difference was observed between the activities of melittin and magainin. Thus, the peptides at non-pore-forming concentrations can form small pores permeable for protons but not for calcein.

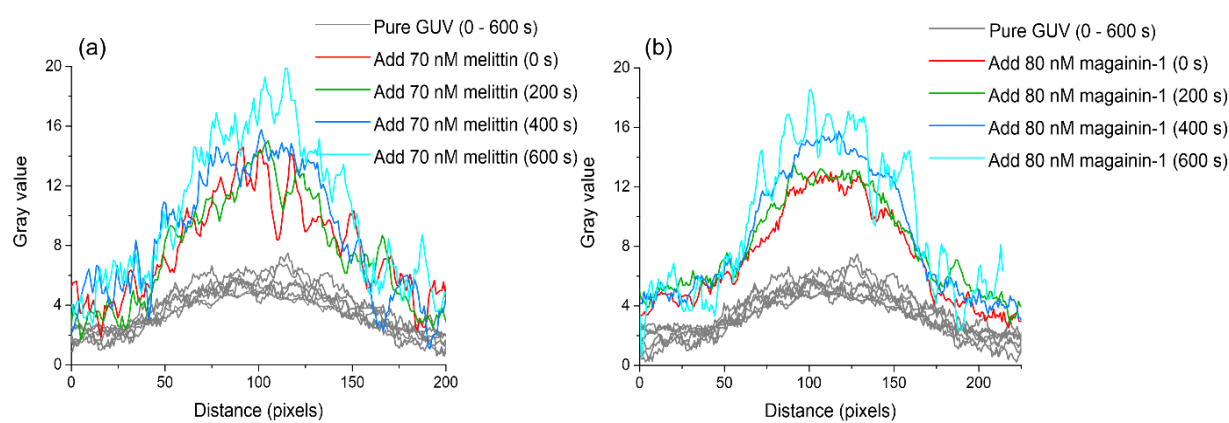

FIG. 8. Densitometry plots of GUVs comprising DOPC:DOPG:DOPE (60:20:20 mol.%) with adsorbed 70 nM melittin (a) and 80 nM magainin (b), introduced through a micropipette in the vicinity of the vesicle. The brightness profiles were plotted along the diameter of the GUVs during 600 s. The fluorescence intensity of intact GUVs in the absence of peptides (dark gray) was constant during 600 s (n = 30). The addition of melittin (n = 50) or magainin (n = 50) caused an increase in the brightness level of the GUV. Measurements were performed each 200 s from the start of peptide addition (0 s) to 600 s.

We also evaluated the possible proton transport through the GUV membrane comprising DOPE:DOPG:TOCL = 70:20:10 mol.%. In order to identify the concentration of pre-adsorption that can provide the complete inhibition effect, we decided to start from the lowest concentrations (35 nM melittin and 40 nM magainin). Adding 35 nM melittin [Fig. 9(a)] and 40 nM magainin [Fig. 9(b)] caused an increase in the calcein fluorescence intensity rapidly after we began adding the peptide, with the steady-state after approximately 600 s. The increase in the fluorescence intensity was consistent with a pH increase inside the GUV. No significant difference was observed between the activities of melittin and magainin. The results on DOPE:DOPG:TOCL (70:20:10 mol.%) are consistent with the previous experiments on the GUV comprising DOPC:DOPG:DOPE (60:20:20 mol.%). However, if for DOPC:DOPG:DOPE we observed almost complete inhibition effect (1.1 ± 0.5 % of calcein leakage) at the concentration of 80 nM (Table IV), similar effect (5.7 ± 4.0% of calcein leakage) was obtained for DOPE:DOPG:TOCL at the lowest concentration of 40 nM magainin (Table IV). Thus, we can conclude that the presence of cardiolipin in the lipid composition enhances the membrane binding of AMP.

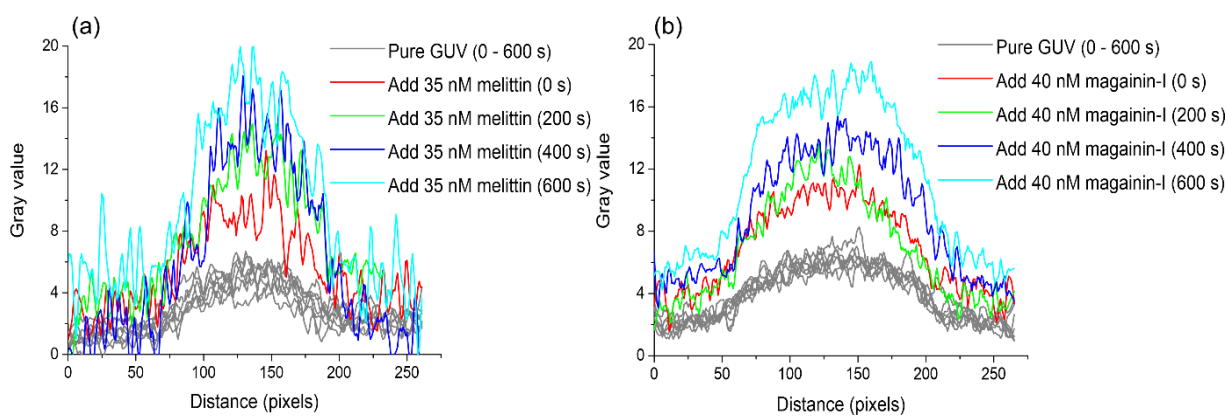

FIG 9. Densitometry plots of GUVs comprising DOPE:DOPG:TOCL = 70:20:10 mol.%. with adsorbed 35 nM melittin (a) and 40 nM magainin (b), introduced through a micropipette in the vicinity of the vesicle. The brightness profiles were plotted along the diameter of the GUVs during 600 s. The fluorescence intensity of intact GUVs in the absence of peptides (dark gray) was constant during 600 s (n = 30). Adding melittin (n = 40) or magainin (n = 40) caused an increase in the brightness level of the GUV. Measurements were performed each 200 s from the start of peptide addition (0 s) to 600 s.

# IV. DISCUSSION

Antimicrobial peptides are the component of the innate immune system of many organisms, from bacteria to humans. For decades, people try to understand why evolution has made a bet on these substances as antimicrobial defense system. The need to reach a detailed understanding in this question becomes more and more urgent because of rapid increase in the number of multidrug-resistant bacteria and simultaneous decrease in our ability to produce new "classical" antibiotics. However, having in a pocket more than 3000 of different antimicrobial peptides, we are still far from understanding their structure-activity relationship. There are several open questions, which should be addressed to reach this goal. First, it is commonly believed that α-helical peptides initially adsorb onto the membrane approximately parallel to its surface (S-configuration) rearranging further into the transmembrane I-configuration, and, finally, translocate through the membrane. Doing so, AMPs form lipid-peptide pores with different structure and lifetime. Despite the differences in structure, charge, and size of the antimicrobial peptides, this scheme is assumed to be common for all of them [47]. Here one can see the first obstacle in developing an effective AMP-based drug. High positive charge of the peptide makes negatively charged surface of bacteria more attractive for them than the neutral one of eukaryotic cells. At the same time, AMPs are amphipathic molecules, and contribution of hydrophobic interactions allows them to adsorb even onto non-charged membranes, although with a lower affinity [86].

The pore-forming concentration of melittin corresponds to a P/L ratio of about 1/25 [72–74]. Taking into account its on-membrane size of 150 $\text{Å}^2$ and the area per lipid headgroup of approximately 70 $\text{Å}^2$ [73], at P/L = 1/25 melittin occupies about 7 % of the membrane surface. For magainin, the pore-forming concentration is P/L = 1/60 [72–74], while it adsorbs onto the membrane in a form of dimers. Thus, we can conclude that it should occupy a comparable part of the membrane surface. Our experimental data (see Table III) suggested that the relative increase of the GUV area was approximately 3%. This is close to the results for magainin-2 and PGLa demonstrating that in GUVs under tension of 0.5 mN/m it is impossible to get an area increase more than 3% even for micromolar concentration of peptides [87,88]. Rough estimations give approximately 3.5% increase of the GUV diameter for the 7% increase in its area, that is close to the experimentally observed values. Note that increase, not decrease, in the GUV area in the pore-forming concentration of different peptides contradicts the idea that peptides just dissolve the membrane by forming peptide-lipid micelles [89].

In the companion paper [62], we suggested that asymmetric adsorption of AMPs onto one monolayer of the membrane leads to an appearance of lateral pressure in this monolayer, while the opposing monolayer should stretch, generating the corresponding lateral tension [Fig. 10, left(b)]. This imbalance creates conditions for the intensive formation of relatively large metastable lipidic pores leaky for ions and large molecules like calcein [Fig. 10, left(c)]. As the lateral pressure/tension are generated in the whole membrane, this mechanism of pore formation can be referred to as the "non-local", as pores can readily form at any point at the membrane, independently on the distance to the nearest peptide molecule. These pores relax the lateral pressure/tension difference in the membrane. When both membrane leaflets are covered by peptides, the imbalance of the pressure/tension disappears, and the peptides adsorption should stop, thus, leading to the pore closure observed after approximately 5-10 minutes of the experiment (Fig. 1).

If peptides in the pore-forming concentration occupy the area *A* in the outer monolayer of the GUV membrane, the lipids in this monolayer should compress by *A*/2, while the lipids in the inner monolayer should stretch by *A*/2; the total change of the GUV membrane area should be +*A*/2 [62]. This agrees well with our experimental data on the change of GUV diameter upon adsorption of melittin or magainin (Table III) and published results for other AMPs [88]. Therefore, amphipathic peptides adsorbing onto one side of the membrane in the S-configuration change its physical properties by adding area to it and producing lateral pressure in the contacting monolayer and lateral tension in the distant monolayer. At some threshold surface concentration of peptides, this results in intensive formation of lipidic pores in a membrane. At this moment, peptides can fall into the membrane and stand parallel, tilted, or in the I-configuration at the edge of the lipidic pore, with possible further translocation to the distant lipid monolayer [Fig. 10, left(d)]. Because pores in a membrane reduce the lateral pressure/tension imbalance, the conductance should shortly disappear. That is why in our experiments and in other papers like [56] one can see that the main contribution to the membrane conductance comes from relatively short-living spikes (see Fig. 2). The size of the pores corresponding to the spikes (Fig. 3) is the same as the size of lipidic pores appearing in the membrane to reduce pressure/tension gradients, as estimated in the companion paper [62]. We assume that the spike conductance corresponds to the lipidic pore [Fig. 2(d)]. Multi-level conductance traces most likely represent peptide-lipid pores in which the peptides dynamically fluctuate so that the tilted peptide molecules can stand at the edge of the pore or escape from it [Fig. 2(c)]. Finally, square-top signals are some long-living stable toroidal pores formed by lipids and several peptide molecules, apparently standing in the I-configuration at the pore rim [Fig. 2(e)]. This configuration rarely appears in the system [53]. Thus, both experimentally and theoretically (see companion paper [62]), we explained that pore formation by α-helical AMPs goes through the non-local modification of the membrane physical properties and the formation of lipidic pores. The process of lipidic pore formation is mostly independent on the peptide structure and determined only by the amphipathicity of the peptide molecule and should be common for a large variety of α-helical antimicrobial peptides.

The rate of lipid flip-flop (passage of individual lipid molecules through the membrane) should increase in membranes with lateral pressure in one monolayer and lateral tension in the opposing monolayer. The passage should be facilitated in the direction from monolayer with pressure to monolayer with tension, as the asymmetry in pressure/tension changes the otherwise equal energies of lipid molecules in two monolayers. When the pressure and tension are equal by the absolute value, the difference in the elastic energy can be estimated as: $\Delta W = 2\sigma a_0$, where $\sigma$ is the value of the tension, $a_0$ is the area per lipid molecule. For typical $\sigma \approx 4$ mN/m, $a_0 = 0.7$ nm$^2$, this yields $\Delta W \approx 1.4$ $k_BT$ ($k_BT \sim 4 \times 10^{-21}$ J). In the most optimal scenario, this difference directly decreases the energy barrier of flip-flop. A typical rate of flip-flop for common phospholipids is about 1 week. Thus, the difference in pressure/tension in lipid monolayers can increase the rate by $\exp[\Delta W/(k_BT)] = \exp(1.4) \approx 4$ times, thus transforming 1 week to about 2 days. Anyway, this time period is much longer than the characteristic time of our experiments. Thus, flip-flop is not expected to play any essential role in releasing pressure differences and excess area.

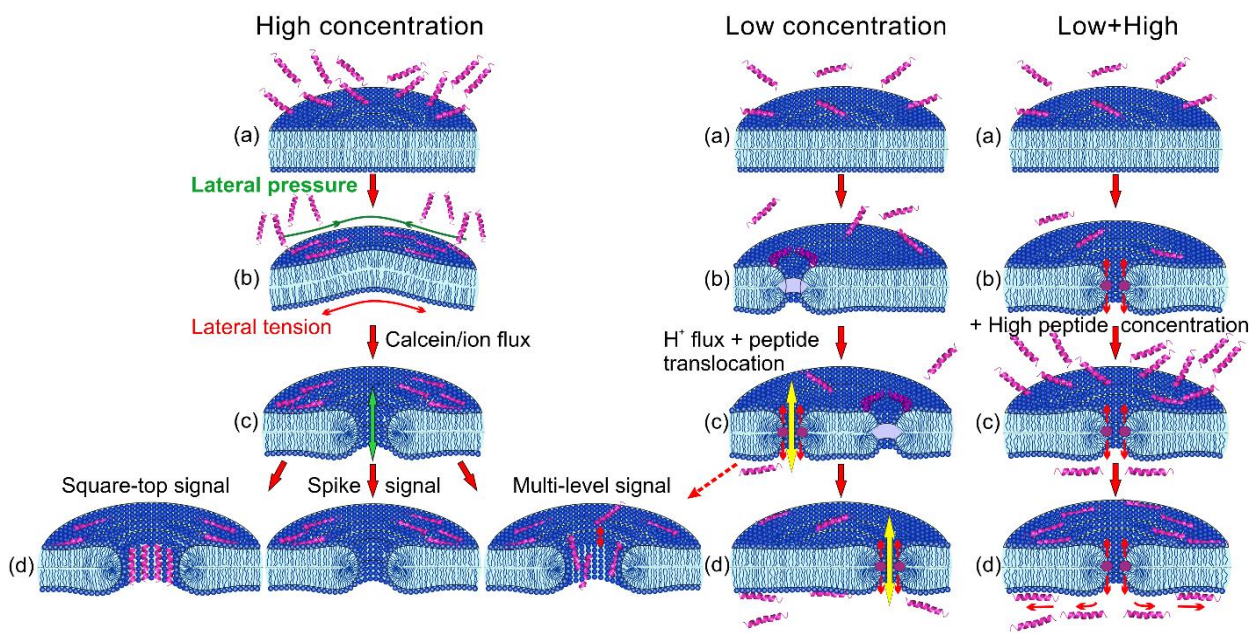

FIG. 10. Scheme of the AMPs membrane activity. High concentration (left): (a) peptides adsorb onto the membrane; (b) peptide adsorption generates an imbalance of the lateral pressure/tension in the opposing membrane leaflets; (c) formation of purely lipidic pores reliefs the lateral pressure/tension; (d) lipidic pore is transiently open (spike signal), expands to the peptide-lipid pore by the peptides standing tilted (multi-level signal) or in the I-configuration at the edge of the pore (square-top signal). The green arrow indicates that a large lipidic pore (similar to the peptide-lipid or toroidal pore) allows ion and calcein flux. Low concentration (middle): (a)

peptides adsorb onto a membrane; (b) two peptides staying at some equilibrium distance form a hydrophobic defect in the membrane between them; (c) hydrophobic defect converts to the hydrophilic $H^+$-pore with the peptides standing at its equatorial plane; (d) at every moment of time peptides form an equilibrium population of these metastable $H^+$-pores in the membrane. The yellow arrow indicates that $H^+$-pore allows proton flux and facilitates peptide translocation. The red dashed arrow indicates that if the surface concentration of the peptides is high enough for the metastable $H^+$-pore to meet the peptides during its lifetime, this $H^+$-pore can expand to a the relatively large peptide-lipid pore inducing multi-level or even square-top signals. Low+High (right): (a) peptides adsorb onto the membrane; (b) two peptides form a hydrophilic $H^+$-pore standing at its equatorial plane (after forming a hydrophobic defect); (c) peptides are added at high (pore-forming) concentration to the membrane with $H^+$-pores; (d) the peptides can escape from the metastable $H^+$-pore onto the opposing side of the membrane, thus, providing a mechanism of intensive peptide transportation across the membrane, even in case of high peptide concentration, without the formation of large pores. Peptide and lipid flow via the population of metastable $H^+$-pores relaxes the otherwise sharp gradient in the lateral pressure/tension between two membrane leaflets. The process of pore formation under the action of AMP in high concentration also includes all the stages of pore formation present in low AMP concentration.

Another open question is the ability of AMPs to block conductance upon pre-adsorption onto the membrane [44]. The authors suggest that the pore-forming activity of peptides disappears after total coverage of the distant membrane leaflet by translocating peptides. However, in our experiments we explicitly took 20- and 40-times lower concentrations of melittin and magainin than those required to form pores. We observed that even in these cases pre-adsorption of the peptides blocks further release of calcein after adding the pore-forming concentration of the peptide (Fig. 7). Moreover, patch-clamp experiments revealed that mainly spike signals disappeared (Table I). Keeping in mind that the spike conductance corresponds to lipidic pores, which diminish the imbalance of pressure/tension between lipid monolayers, we have been surprised by this result. In experiments with calcein release from GUVs, we detected that one-side additions of low concentration of magainin gave twice higher levels of leakage in the presence of pore-forming concentration of the peptide compared to the case of two-side addition. And, finally, peptide pre-adsorption from the inside of the GUV gave the same result as from the outside of the vesicle, which refutes the idea of electrostatic repulsion between adsorbed and incoming peptide molecule as the crucial factor of inhibition of pore formation. Therefore, we can hypothesize that if peptides in a low concentration are pre-adsorbed, it should allow their translocation across the membrane without ion and calcein conducting pores, and should block further formation of such pores in the presence of high, pore-forming concentrations of the peptide. Alternatively, we can suggest that this protective effect depends only on total concentration of the peptides in both monolayers and, possibly, has a threshold concentration value.

In [90] authors observe graded increase of the vesicle leakage upon addition of the GALA peptide in very low concentrations. The leakage of the fluorescent dye has been detected in all of their experiments starting from the lowest reported concentration of 12 nM of the peptide. However, most probably, this finding is an experimental artifact caused by improper preparation of the vesicles and incorrect procedure of determination of the extent of the dye leakage. The vesicles in [90] were prepared by extrusion through porous membrane one or three times, while normally about 10 times of extrusion are necessary (e.g., in the work [91] the vesicles were extruded 10 times); this, most likely, resulted in a large population of multilamellar vesicles. In [90], the vesicle suspension passed a single step of freeze-thaw, while normally 5 such steps should be done, e.g., as in [91]. The fluorescence measurements to determine the leakage mechanism used the fluorophore/quencher pair of ANTS/DPX. After the incubation with the peptide GALA, the vesicles were separated from the leaked dye and quencher in a column, and their fluorescence was determined after that. This approach strongly relies on the fact that GALA-treated vesicles do not rupture in the column. Note that in [92] authors have not detected GUV leakage even for 0.5 μM of the GALA peptide in solution for similar membrane lipid composition. Therefore, real pore-forming concentration of the GALA peptide might be higher than those manifested in [90], and, to the best of our knowledge, nobody has tried to study the protective effect of the none-pore-forming concentrations of the peptides.

Would the protective effect of the peptides be possible without formation of any defect in the membrane? To remove this contradiction, we performed experiments with proton transport across the membrane in the presence of low non-pore-forming concentrations of melittin and magainin. Our data clearly demonstrated that in these low concentrations GUV membrane became permeable for protons. Recently, we developed a full trajectory of lipidic pore formation in a membrane, which starts from an intact bilayer, passes via the so-called hydrophobic defect with the ultimate formation of a hydrophilic pore [93]. The states of the intact bilayer and the hydrophilic pore are separated by the energy barrier, whose height depends on the membrane elastic properties. In the companion theoretical paper [62] we showed that two peptide molecules decreased the energy of the hydrophilic pore formation [Fig. 10, middle(b)] in the membrane between them, if they stayed at some equilibrium distance [36]. In this configuration, shallowly inserted amphipathic peptides line the inner lumen of the $H^+$-pore standing at its equatorial plane, still being parallel to the membrane [Fig. 10, middle(c)]. This $H^+$-pore makes an incessant contact between the opposing membrane leaflet. In other words, two peptides form an $H^+$-pore, or bridge between the outer and inner lipid monolayers preventing further appearance of lateral pressure/tension imbalance between membrane leaflets due to lipid and peptide flow. The size of this $H^+$-pore is too small to conduct water-soluble fluorescent dyes or even ions. However, they are permeable for protons and, possibly, water. The calculations in the companion paper [62] predict that the $H^+$-pores should be metastable, with finite average lifetime. Thus, at every moment of time we have some equilibrium population of these $H^+$-pore in a membrane. The peptides tend to be on the equator of the $H^+$-pores. From such configuration peptides with equal probability may translocate to any of the membrane leaflets. The $H^+$-pores thus allow transportation of incoming peptides across the membrane, even in the case of high peptide concentration, without appearance of the pressure/tension gradients (Fig. 10, right). If the surface concentration of the peptides is high enough for a metastable $H^+$-pore to meet a peptide molecule during its lifetime, the peptide molecule can be captured by the $H^+$-pore [Fig. 10, left(d), middle(c)]. In this case, the $H^+$-pore radius should increase, as well as its lifetime, as the $H^+$-pore comprising three peptide molecules is more stable than the one comprising two [94]. Thus, under certain conditions, relatively large peptide-lipid pores can be formed by a local mechanism [62] even in the absence of lateral pressure/tension gradients. The presence of rare long-living conducting events in patch-clamp experiments with pre-adsorption may be explained in this line.

One can see that magainin, in contrast to melittin, at some concentration could fully block calcein release. Also, there is some discrepancy between the effectiveness of this process in case of GUVs and in patch-clamp experiments. In the companion paper [62], we predicted that the difference in local membrane activity of single amphipathic peptides is dictated by the degree of their immersing into the lipid monolayer, parameterized by the projection of the boundary director $|n_0|$. Because melittin disturbs the membrane to a less extent compared to the dimers of magainin [70,95], the energy barrier for transition from an intact bilayer to hydrophilic $H^+$-pore in the presence of melittin is higher than in the presence magainin; on the contrary, the energy barrier of the $H^+$-pore closure for melittin is smaller than for magainin. Altogether, this means that the lifetime of the $H^+$-pore is shorter, and the equilibrium concentration of the $H^+$-pores is smaller for melittin as compared to magainin. This may be the reason of the incomplete protection of the membrane pre-treated with a low concentration of melittin: a small number of $H^+$-pores is unable to effectively relief the gradient of lateral pressure/tension increasing upon the subsequent treatment of the membrane with a high concentration of melittin. Additionally, the theoretical model predicts that lateral tension of the membrane will generally increase the surface concentration of AMPs at the same bulk concentration [79]. It is known that the tension of membranes of free GUVs is very low, if exists, while in patch-clamp experiments it is much higher [67]. Therefore, we twice decreased the pre-adsorption concentration of the peptides and obtained almost complete inhibition of pore formation in the case of magainin (Tables II, III), while nothing changed for melittin. These results support the predictions of our theoretical model [62].

What is the consequence of these local effects of the low concentration of peptides? They suggest that a decrease in the AMP concentration to the nanomolar range protects the membrane from large pore formation upon further addition of much higher amounts of peptides. Thus, trying to reduce side-effects of AMPs by decreasing their working concentration, one protects the bacteria instead of killing them, and helps them gain resistivity. On the other hand, pre-treatment of eukaryotic cells and "good" bacterial membranes with some amphipathic molecules close in their structure to AMPs, may aid their defense against treatment with high concentrations of antimicrobial peptides. However, this could be affected by the loss of the asymmetry of lipid composition between membrane leaflets.

To summarize, we can say that in the two companion papers we provided a comprehensive theoretical and experimental study of all possible ways of membrane activity of α-helical amphipathic AMPs. We showed that, depending on the concentration, peptides can locally and non-locally change the membrane structure and its permeability. Even in nanomolar concentrations AMPs make the membrane permeable to water and protons, while this effect protects the membrane from increased amounts of peptide rather than destroying the membrane. At high initial concentration amphipathic peptides provide an imbalance of lateral pressure/tension between membrane leaflets. This imbalance results in purely lipidic pores allowing lipid translocation. Therefore, all witnesses of AMP membrane activity may be described in a framework of one physical model, based on added area, lipidic pores, $H^+$-pores and peptide-lipid pores. The conclusions of these studies provide us with necessary information for the development of an effective AMP. First, we need to stabilize the long-living channel-like structures by specific interactions between peptides. Second, lowering the AMP concentration protects the membrane and may help bacteria gain resistivity. And third, the same effect one can use to pre-treat healthy and necessary cells from therapy with high concentrations of AMPs.

## ACKNOWLEDGMENTS
This research was funded by the Ministry of Science and Higher Education of the Russian Federation.

## DATA AVAILABILITY STATEMENT
The data that support the findings of this study are available from the corresponding author upon reasonable request.

## AUTHOR DECLARATIONS
### Conflict of Interest
The authors have no conflicts to disclose.